\documentclass[%
 reprint,
 amsmath,amssymb,
 aps,
]{revtex4-2}

\usepackage{graphicx}
\usepackage{dcolumn}
\usepackage{bm}
\usepackage{multirow}
\usepackage{tabularx}
\usepackage{booktabs}
\usepackage{makecell}
\usepackage{float}

\begin{document}

\preprint{APS/123-QED}

\title{Highly nondegenerate polarization-entangled photon pairs produced through noncritical phasematching in single-domain KTiOPO$_4$ }
\author{Jia Boon Chin$^1$}
 \email{jiaboonc@gmail.com}
\author{Diane Prato$^1$}%
\author{Alexander Ling$^{1,2}$}
\affiliation{
 $^1$Centre for Quantum Technologies, National University of Singapore, 3 Science Drive 2, Singapore 117543, Singapore\\
 $^2$Physics Department, National University of Singapore, Faculty of Science, 2 Science Drive 3, Blk S12, Level 2, Singapore 117551.}%


\begin{abstract}

Photon-pair sources are useful for entanglement distribution. The most mature of these are spontaneous parametric downconversion (SPDC) sources, most of which achieve phasematching via engineering the domains in poled crystals or the angle between the optic axis and the pump beam. 
For multi-channel entanglement distribution of photon pairs, where one photon is transmitted through free-space and the other photon is transmitted through fiber, it is beneficial to use highly nondegenerate photon-pair sources. The currently accepted approach in such sources is quasi-phasematching. In this paper, a simpler, more stable alternative is presented for producing highly nondegenerate photon pairs. A source of polarization-entangled photon pairs with low temperature sensitivity based on noncritical phasematched (NCPM) SPDC in single-domain potassium titanyl phosphate (KTP) was demonstrated. Over a crystal temperature range of $75^\circ$C, the center wavelength of the idler photons was observed to change by $10.8$nm while 
the average entanglement visibility was maintained above 98\%. With the signal photons detected locally, the idler photons were transmitted through 62km and 93km of deployed telecom fibers with average raw visibilities of $98.2(1)\%$ and $95.6(3)\%$ respectively. 
 
\end{abstract}

\maketitle

\section{\label{section:intro}Introduction}

Entanglement is an important resource in quantum networks \cite{Wehner2018}. Entangled photon pairs in particular are useful for sensing, secure communication and computation \cite{Prevedel2007, Gisin2007, O’Brien2009}. They can serve as interfaces between matter-qubits and can be used to distribute entanglement to realize a widespread quantum network.

The current workhorse for generating entangled photon pairs is through spontaneous parametric downconversion (SPDC). Various source designs have been used to generate polarization-entangled photon pairs \cite{Anwar2021}. There are two main approaches to phasematching used in these sources. Most designs in the past decade use quasi-phasematching (QPM) achieved via the modulation of the nonlinear coefficient $\chi^{(2)}$ in the downconversion crystal. These poled crystals, typically lithium niobate (LN) or potassium titanyl phosphate (KTP), are customized in their poling periods to achieve QPM and generate SPDC photons at target wavelengths for a selected pump wavelength and operating temperature. As the phasematching conditions are sensitive to temperature, these sources require temperature regulation typically to within $0.1^\circ$C \cite{Szlachetka2023, Jabir2017, Steinlechner2014}. 

The second approach is critical phasematching (CPM), where the cut angle of the crystal is chosen and the phasematched SPDC wavelengths are tuned by varying the angle between the optic axis and the pump beam. This angle causes spatial walkoff between the extraordinary and ordinary beams, although the effect of this can be mitigated by using compensation crystals \cite{Anwar2021, Villar2018, Trojek2008}.

There is a third and less common approach known as noncritical phasematching (NCPM). In NCPM, the interacting waves are polarized along the axes of the crystal. The phasematching criteria for co-propagating NCPM SPDC is given by
\begin{eqnarray}
\Delta k = k_{p} - k_{s} - k_{i} = 2\pi\left(\frac{n_p}{\lambda_p} - \frac{n_s}{\lambda_s} - \frac{n_i}{\lambda_i}\right) = 0,
\label{eq:phasematching}
\end{eqnarray}
where $k_j$ refers to the wavevector magnitude of the pump, signal or idler.
The lack of spatial walkoff effects in NCPM simplifies photon-pair collection and allows the use of longer crystals which can increase the pair generation rate and reduce the output bandwidth \cite{Fedrizzi2007, Bennink2010}. Since poling is not required, the downconversion crystals are less costly and can be made longer with wider apertures.

SPDC photon-pair sources are typically designed for transmission of the signal and idler photons over the same channel or for detection with the same optics. They generally target degenerate or near-degenerate output in the visible (VIS) to near-infrared (NIR) wavelengths \cite{Pseiner2021, Perumangatt2020, SansaPerna2022, Yin2017, Anwar2022, Mishra2022, Omshankar2024}, or in the O- and C-band wavelengths \cite{Shi2020, Wengerowsky2018, Wengerowsky2020, Aktas2016}. However, degenerate or near-degenerate output in NCPM SPDC typically lie between these two wavelength regimes for the nonlinear crystals commonly used in these sources. For example, degenerate NCPM SPDC in KTP has been reported at wavelengths around 1080nm \cite{Kim2024}. Furthermore, the range of output wavelengths available from NCPM are relatively inflexible because NCPM has neither the customizability of the poling period nor the tunability of the crystal angle available to the other approaches. These make NCPM an unpopular approach in SPDC photon-pair sources.

\begin{figure*}[ht]
  \centering
  \includegraphics[width=1\textwidth]{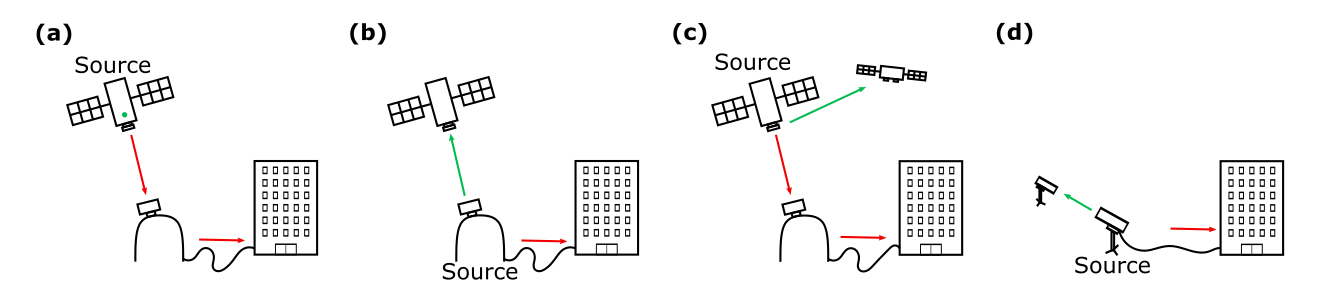}
\caption{Use cases of VIS/NIR-telecom photon pairs. The VIS/NIR signal photons are denoted by green and the telecom idler photons are denoted by red. (a) Signal photons detected onboard and idler downlink, assuming use of adaptive optics; (b) Signal uplink; (c) Signal photons used for inter-satellite communication and idler downlink, assuming use of adaptive optics; (d) Signal photons used to communicate with terrestrial free-space terminal. The idler photons can be transmitted through optical fibers with low attenuation. The signal photons can be efficiently detected by compact and portable Si-APDs and are suited for free-space transmission as there are less diffraction effects.}
\label{fig:usecases}
\end{figure*}

Despite the drawbacks, NCPM has its advantages in some use cases. NCPM readily provides highly nondegenerate SPDC output where the signal and idler photons have very different wavelengths, which can be beneficial for quantum sensing and imaging \cite{Chen2019, Lualdi2022, Pearce2023}. 

Of particular interest for quantum communication are cases where the signal wavelength lies in the VIS to NIR band, while the idler wavelength lies in the telecommunication bands with low attenuation in optical fibers. 
Currently, compact and portable single-photon detectors with high detection efficiencies in the telecommunication wavelengths are unavailable. In contrast, commercial silicon avalanche photodiodes (Si-APDs) with small footprints and high detection efficiencies in the VIS/NIR wavelengths are mature enough to be deployed on satellites. 
It is thus practical for quantum network nodes with tight size, weight, and power (SWaP) requirements to be equipped with Si-APDs and communicate in VIS/NIR bands. On the other hand, the nodes connected to the extensive telecommunication infrastructure will likely favour communication at the telecommunication wavelengths due to the low attenuation loss in fiber and the availability of superconducting nanowire single-photon detectors (SNSPDs) that have near-unity detection efficiencies. VIS/NIR-telecom photon-pair sources can serve as interfaces between these two types of nodes through multi-channel entanglement distribution. Figure \ref{fig:usecases} depicts some of these scenarios. 

In the context of satellite quantum communication, a satellite equipped with such a highly nondegenerate entangled photon-pair source can detect the signal photon onboard with Si-APDs while sending the idler photon to a ground station in a downlink \cite{Bedington2017}. If the ground station is equipped with adaptive optics so that the idler photons can be coupled into standard telecommunication fibers, quantum communication between the satellite and locations connected to the ground station via fiber can be enabled (Fig. \ref{fig:usecases}(a)).  

The signal photons need not be detected locally at the source. Leveraging on the portability of Si-APDs and relatively low diffraction of light at visible wavelengths compared to longer wavelengths, the signal photons are well-suited to be sent through free-space to be detected by mobile nodes. The signal photons can be sent from ground to satellite as an uplink (Fig. \ref{fig:usecases}(b)) or between satellites (Fig. \ref{fig:usecases}(c)). In a metropolitan setting, an entangled photon-pair source on the ground can also link portable free-space terminals to areas with access to the fiber network (Fig. \ref{fig:usecases}(d)). Drones or vehicles could also be interfaced as mobile nodes in more complex and mature quantum networks \cite{Liu2020, Liu2021, Bourgoin2015}.

Some examples of these VIS/NIR-telecom photon pairs achievable through NCPM SPDC in nonlinear crystals are shown in Table \ref{tab:ncpmwavelengths}. Depending on the crystal, different pump wavelengths are required to achieve the idler wavelengths in the O-band or C-band. While some pump wavelengths may not be readily available, multi-channel entanglement distribution is still viable if the idler wavelength lies within a telecom band. The effective nonlinearity coefficients of the NCPM processes are generally lower than those used in Type-0 QPM SPDC sources, resulting in lower pair generation rates in NCPM SPDC. However, NCPM SPDC sources using KNbO$_3$ crystals can be expected to be as bright as Type-0 QPM SPDC sources due to the high nonlinearity.

In this work, we demonstrate a source of highly nondegenerate polarization-entangled photon pairs using the NCPM approach. We chose KTP as the downconversion crystal due to its widespread availability, low cost, and its high temperature-stability in the NCPM process compared to the QPM processes as elaborated in the next section.

\begin{table*}[ht]

\begin{ruledtabular}
\caption{\label{tab:ncpmwavelengths}Potentially useful wavelength triplets from NCPM SPDC in various nonlinear crystals. $\lambda_s, \lambda_i$ refers to the expected NCPM output signal and idler wavelengths when a pump of wavelength $\lambda_p$ is used. The letters in brackets denote the axes of the crystal along which the waves are polarized. KNbO$_3$ crystals have high $d_{\text{eff}}$ for the listed NCPM processes, suggesting that NCPM SPDC sources that utilize these crystals can be comparable in brightness to those based on Type-0 $z_pz_sz_i$ or $e_pe_se_i$ QPM processes in KTP and LN.}
\begin{tabular}{ccccccc}

Crystal&$\lambda_{p}$(nm) & $\lambda_{s}$(nm) & $\lambda_{i} (nm)$ & Temperature($^\circ$C) & \thead{Nonlinear coefficient \\ (assuming Kleinmann \\ symmetry condition)} & \thead{$|d (1.064\mu m)|$ \cite{Nikogosyan2006}\\(pm/V)} \\ \hline
 \multirow{2}{5em}{KTiOPO$_4$}
    &   $443(y)$    &   $668(z)$    &   $1310(y)$   &   30  &  $d_{32}$ &   3.7  \\
    &   $387(y)$    &   $516(z)$    &   $1550(y)$   &   30  &  $d_{32}$ &   3.7  \\ \addlinespace

 \multirow{4}{5em}{RbTiOPO$_4$}
    &   $402(x)$    &   $581(z)$    &   $1312(x)$   &   30  &  $d_{31}$ &   3.3  \\
    &   $345(x)$    &   $444(z)$    &   $1551(x)$   &   30  &  $d_{31}$ &   3.3  \\
    &   $493(y)$    &   $790(z)$    &   $1310(y)$   &   30  &  $d_{32}$ &   4.1  \\
    &   $426(y)$    &   $587(z)$    &   $1552(y)$   &   30  &  $d_{32}$ &   4.1  \\ \addlinespace
    
 \multirow{4}{4em}{KNbO$_3$}    
    &   $405(z)$    &   $586(x)$    &   $1311(x)$   &   58  &  $d_{31}$ &   11.9  \\
    &   $375(z)$    &   $495(x)$    &   $1550(x)$   &   40  &  $d_{31}$ &   11.9  \\
    &   $488(z)$    &   $778(y)$    &   $1310(y)$   &   83  &  $d_{32}$ &   13.7  \\
    &   $461(z)$    &   $656(y)$    &   $1550(y)$   &   62  &  $d_{32}$ &   13.7  \\ \addlinespace

 \multirow{2}{6em}{5\%MgO:LiNbO$_3$}
    &   $515(e)$    &   $846(o)$    &   $1316(o)$   &   45  &  $d_{31}$ &   4.4  \\
    &   $515(e)$    &   $771(o)$    &   $1551(o)$   &   72  &  $d_{31}$ &   4.4 \\ 
\end{tabular}
\end{ruledtabular}

\end{table*}

\section{\label{section:NCPMinKTP}Noncritical phasematching in KTP}

\begin{figure}[ht]
  \centering
  \includegraphics[width=0.5\textwidth]{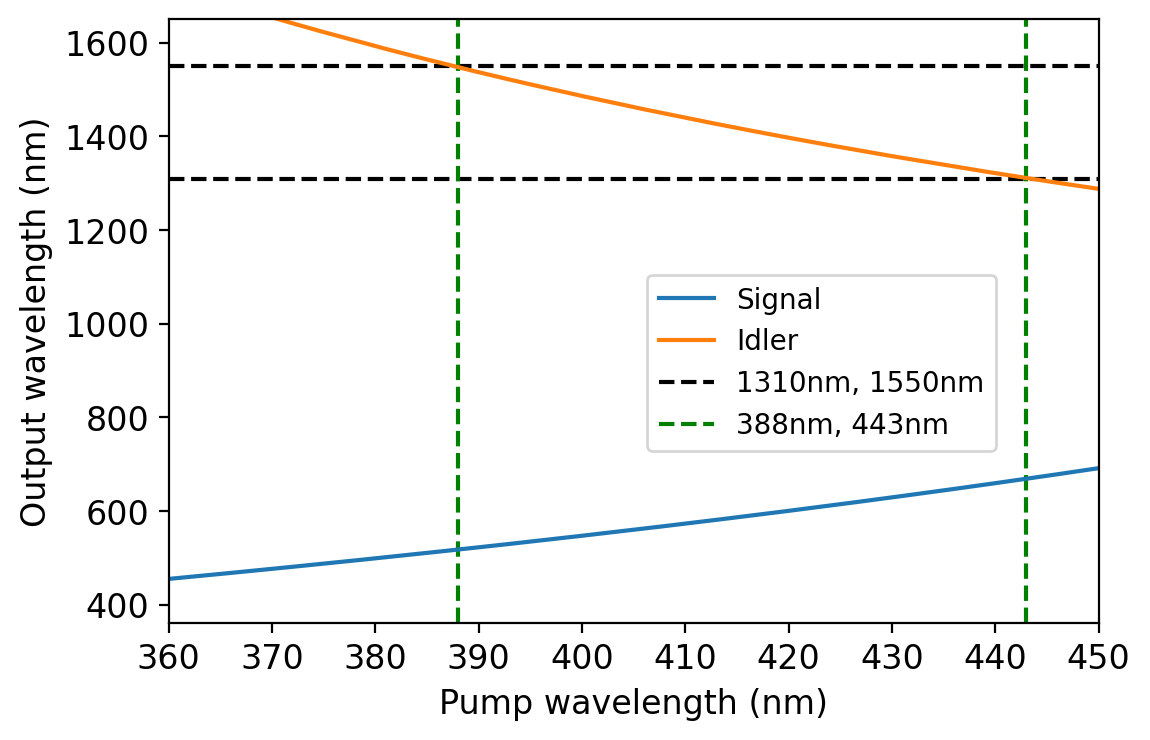}
\caption{Calculated wavelengths of output from NCPM SPDC in KTP at T = 60$^\circ$C. The pump and idler are y-polarized and the signal is z-polarized.}
\label{fig:KTPyzy}
\end{figure}

In KTP, visible-telecom photon pairs from NCPM SPDC can be achieved by propagating a y-polarized pump in the 360nm - 450nm range along the crystal's x-axis (see Fig. \ref{fig:KTPyzy}). The downconverted visible- and telecom-wavelength photons are z- and y-polarized, respectively.

\begin{figure}[ht!]
  \centering
  \includegraphics[width=0.5\textwidth]{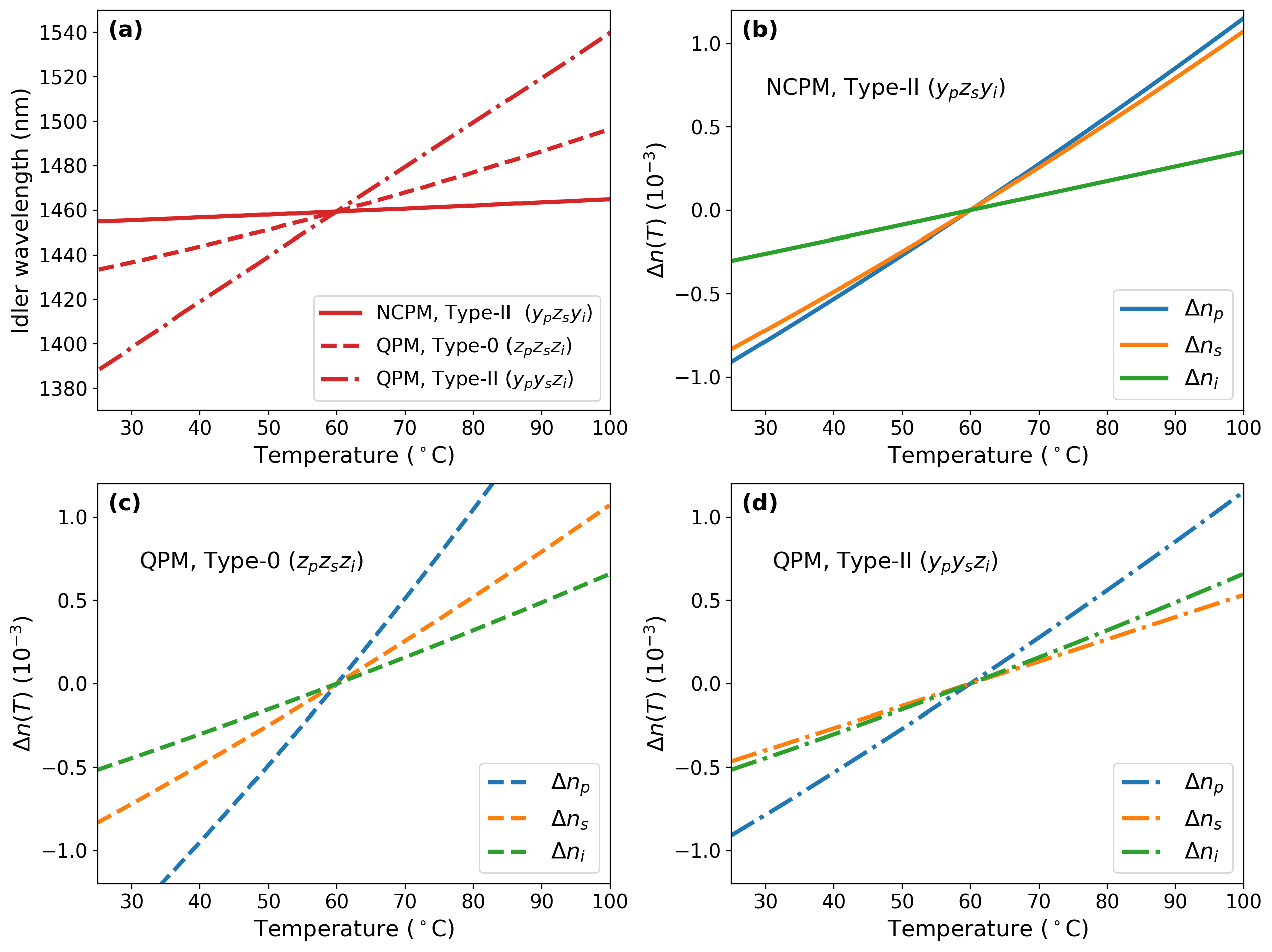}
\caption{(a) Calculated output idler wavelengths from NCPM, Type-0 QPM and Type-II QPM SPDC processes vs temperature of KTP. Variation of refractive indices for the pump, signal and idler with temperature,  from a reference of temperature of T = 60$^\circ$C for (b) NCPM, (c) Type-0 QPM and (d) Type-II QPM SPDC processes. The poling periods for the QPM processes were chosen to produce the same wavelengths as NCPM at T = 60$^\circ$C.}
\label{fig:tempanddeltan}
\end{figure}

Numerical calculations from the temperature-dependent Sellmeier equations for KTP \footnote{The Sellmeier equations used for the y- and z- refractive indices were from \cite{Fan1987} and \cite{Fradkin1999} respectively, with their temperature dependence given by \cite{Emanueli2003}.} show that the wavelengths of visible-telecom photon pairs exhibit little sensitivity to temperature changes for the NCPM process compared to QPM processes involving the same wavelengths (Fig. \ref{fig:tempanddeltan}(a)). Over the range of 25$^\circ$C to 100$^\circ$C, the center wavelengths of the idler from the Type-II and Type-0 QPM processes exhibited changes of 2.03nm/$^\circ$C and 0.84nm/$^\circ$C respectively, while the center wavelength of the idler from the NCPM process changes at a rate of 0.13nm/$^\circ$C. This reduced-temperature sensitivity of NCPM is beneficial for stable operation in situations where temperature regulation requires additional engineering effort to meet SWaP constraints, such as on satellites \cite{Chandrasekara2015, Bedington2016} and drones. 

An intuitive reasoning for the low temperature sensitivity can be obtained by rewriting the phasematching equation (\ref{eq:phasematching}) as $n_p - n_s \frac{\omega_s}{\omega_p} - n_i\frac{\omega_i}{\omega_p} = 0$. In an extreme scenario where the output wavelengths do not change with the temperature of the downconversion crystal given a constant pump wavelength, the sum of refractive indices of the signal and idler photons, weighted by the ratio of their respective energies to the pump photon, has to remain equal to the refractive index of the pump in the presence of a temperature change. In light of this, the reduced-temperature sensitivity for NCPM SPDC in KTP in this regime of high nondegeneracy arises due to two factors. 

First, when varying the temperature, the z-refractive index at the signal wavelength changes at almost the same rate as the y-refractive index at the pump wavelength (Fig. \ref{fig:tempanddeltan}(b)). Second, the wavevector of the signal is more dominant than the idler's because of the shorter wavelength ($k_s = 21.0 \mu m^{-1}, k_i = 7.5 \mu m^{-1}$). These two factors mean that the change in $n_p$ due to a temperature change is largely compensated by a similar change in $n_s$, leaving only a small phasemismatch due to a relatively undercompensating change in $n_i$. As a result, the phasematching condition in equation (\ref{eq:phasematching}) can be restored with only a small change in output wavelengths in the presence of a temperature perturbation. In the two QPM processes, the first factor is not applicable and the changes in both $n_s$ and $n_i$ are insufficient (Fig. \ref{fig:tempanddeltan}(c),(d)), resulting in a larger change of output wavelengths required to restore phasematching. 

It is worth noting that QPM processes are not inherently less temperature-stable than NCPM processes despite the effects of thermal expansion on the poling period. Near-degenerate QPM Type-II (also $y_pz_sy_i$) in KTP has been shown to be temperature-stable at the telecom regime, demonstrating a wavelength change of 0.18nm/$^\circ$C, with an even lower temperature tuning rate predicted for degenerate output at 1165nm \cite{Pan2021}. The presence of thermal expansion on the poling period in KTP is not significant compared to the variation in refractive indices for QPM processes. For example, for a temperature change of 1$^\circ$C in typical first-order QPM scenarios, the magnitude of the change in the wavevector contribution from the expansion of poling period is on the order of $10^{-5}\sim10^{-6} \mu m^{-1}$, while that from the variation in refractive index is on the order of $10^{-4}\sim10^{-5} \mu m^{-1}$ per interacting wave.

In addition to robustness to temperature changes, NCPM in KTP can also provide spectrally narrow output. The linewidths for the NCPM and QPM processes at KTP temperature of 60$^\circ$C are shown in Table \ref{tab:linewidthcomparison}, assuming a planewave pump at 405.75nm is used and the length of KTP is 30mm.

Having a narrow linewidth is beneficial in numerous applications for quantum communication.
\begin{table}[hb!]
\label{tab:linewidthcomparison}
\begin{ruledtabular}
\caption{Comparison of output spectrum linewidths from NCPM, Type-0 QPM and Type-II QPM processes from a model assuming a pump wavelength of 405.75nm and a x-cut KTP length of 30mm at $T = 60^\circ$C was used. The poling periods for the QPM processes were chosen to produce the same wavelengths as NCPM at T = 60$^\circ$C. The signal center wavelength is 562.0nm and the idler center wavelength is 1459.4nm. }
\begin{tabular}{p{0.4\linewidth} >{\centering\arraybackslash}p{0.25\linewidth} >{\centering\arraybackslash}p{0.25\linewidth}}
    & Signal FWHM (nm) & Idler FWHM (nm) \\ \hline
 
    NCPM, Type-II ($y_pz_sy_i$)   & 0.04  & 0.24 \\
    
    QPM, Type-0 ($z_pz_sz_i$)   & 0.05  & 0.35 \\

    QPM, Type-II ($y_py_sz_i$) & 0.26 & 1.75 \\
\end{tabular}
\end{ruledtabular}

\end{table}
For example, in free-space links, the narrow linewidth of the signal allows one to aggressively spectrally filter out much of the background noise with little compromise to the signal collection. This can especially benefit daylight QKD or other free-space quantum communication applications in the presence of background light. The narrow linewidth of the telecom idler also means less polarization mode dispersion and chromatic dispersion effects, allowing entanglement distribution through fiber with lower quantum bit error rates \cite{Du2024, Rodimin2024, Zhang2024a, Lim2016}.




\begin{figure*}[ht!]
  \centering
  \includegraphics[width=0.8\textwidth]{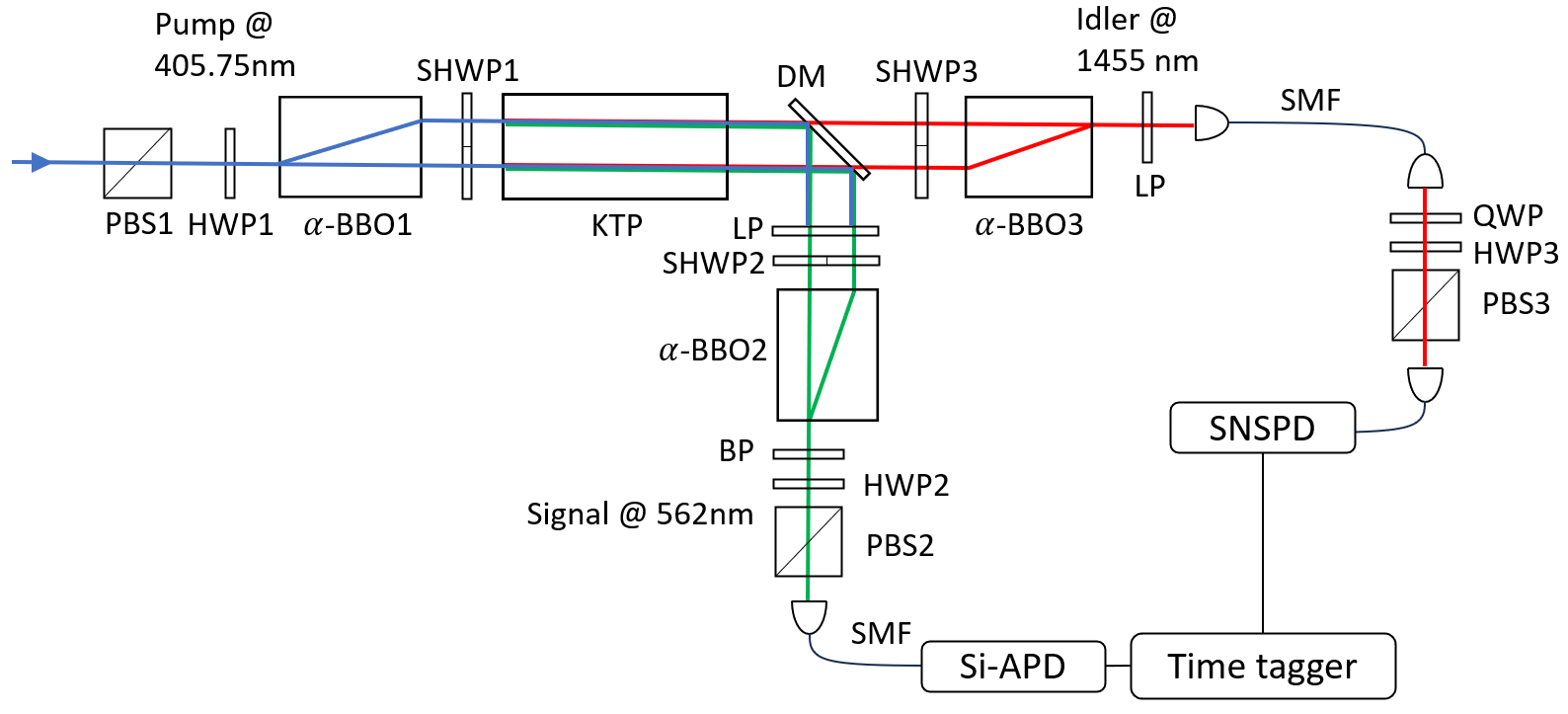}
\caption{Schematic of the setup. PBS: polarizing beam splitter; HWP: halfwave plate; QWP: quarterwave plate; $\alpha$-BBO: alpha-barium borate; SHWP: segmented halfwaveplate, where one segment rotates the polarization of input light by 90$^\circ$ and the other does no rotation; KTP: potassium titanyl phosphate; DM: dichroic mirror; LP: longpass filter; BP: bandpass filter; SMF: single mode fiber; Si-APD: silicon avalanche photodiode; SNSPD: superconducting nanowire single-photon detector.}
\label{fig:schematics}
\end{figure*}

\section{\label{section:source} Polarization-entangled photon-pair source characterization}

To verify the characteristics of NCPM outlined in section \ref{section:NCPMinKTP}, a polarization-entangled photon-pair source based on Type-II NCPM was demonstrated. Polarization entanglement was achieved using a parallel-path interferometric design \cite{Fiorentino2008, Shen2022} with a 50mm single-domain x-cut uncoated KTP as the downconversion crystal. A schematic of the experimental setup is shown in Fig. \ref{fig:schematics}. In this source, a continuous wave pump of wavelength 405.75nm was used. The idler photons lie in the E- and S-bands, and can be transmitted through telecom fibers with low attenuation.

The pump light is injected into the setup and passes through polarizing beamsplitter 1 (PBS1) and halfwave plate 1 (HWP1) to become diagonally polarized. An alpha-barium borate ($\alpha$-BBO1) crystal splits the pump beam into two paths of equal power, with the straight-through path polarized vertically and the laterally displaced path polarized horizontally. The pump photons in the straight-through path have their polarizations rotated by 90$^\circ$ by one segment of segmented halfwave plate 1 (SHWP1), while those from the other path are also delayed by another segment of the same thickness without polarization rotation so that both beams entering the KTP crystal are horizontally polarized. 

The single-domain KTP is oriented with its z-axis vertical and the y-axis horizontal. The NCPM downconversion generates signal photons with wavelength of 562nm and the idler photons with wavelength of 1455nm at room temperature. The pump waist ($w_p\sim110um$) and collection waists ($w_s\sim100um, w_i\sim100um$) were located at the center of the crystal. A dichroic mirror (DM) with an edge of 925nm transmits the idler photons and reflects the signal photons. The pump photons are removed using longpass (LP) and bandpass (BP) filters. 

\begin{figure}[ht]
  \centering
  \includegraphics[width=0.45\textwidth]{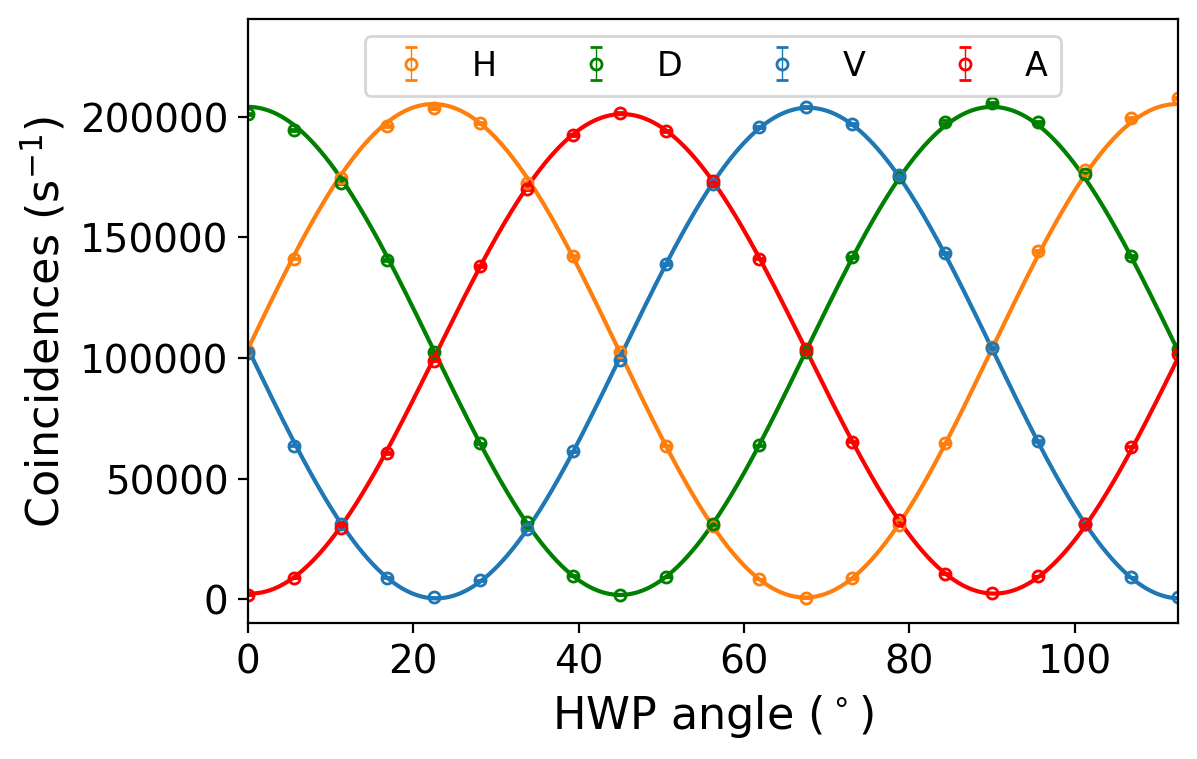}
\caption{Visibility curves for the source with a pump power of $19$mW, before subtracting accidentals. With a coincidence window of $1$ns, there are approximately $550$ accidental pair events per second.}
\label{fig:sourcesweep}
\end{figure}

The signal photons from the two paths have orthogonal polarizations after passing through SHWP2. $\alpha$-BBO2 recombines the two paths to remove distinguishing which-path information. SWHP3 and $\alpha$-BBO3 do the same for the idler photons. The relative phase between the two downconversion paths is set to be an odd multiple of $\pi$ to generate the $|\Phi^-\rangle$ Bell state by tilting $\alpha$-BBO1.

The downconverted photons are collected into single mode fibers (SMFs). The signal photons are detected with a Si-APD while the idler photons are detected with an SNSPD optimized for detection in the C-band. Polarization correlations are analyzed using HWP2, PBS2, HWP3 and PBS3. The quarter waveplate (QWP) corrects for the fiber birefringence in the idler arm before polarization correlation measurements.

The entanglement visibility was measured by sweeping the idler HWP with the signal HWP set to transmit H, D, V and A polarizations (Fig. \ref{fig:sourcesweep}). A coincidence window of 1ns was used, limited by the timing jitter of Si-APD. With 19mW of pump power, approximately $200000 \text{pairs/s}$ were observed at the peak of the coincidence curves (observed brightness of $>10000 \text{pairs/s/mW}$ per polarization mode) with average raw heralding efficiencies of $\eta_s = 25.6\%$ and $\eta_i = 27.4\%$. The visibilities before subtracting for accidental pair events were observed to be $99.33(3)\%, 98.29(4)\%, 99.69(2)\%, 97.74(5)\%$ when the signal was projected to the H, D, V and A polarizations respectively. This resulted in an average raw visibility of $98.77(2)\%$. A CHSH S-parameter of $2.793(3)$ and fidelity \cite{Anwar2021,Chang2016} of $F\geq 0.988$ to $|\Phi^-\rangle$ was obtained from the fit. After correcting for approximately 550 accidental pair events per second, the average visibility was $99.28(1)\%$, with $S = 2.807(3)$ and $F\geq 0.993$. Through time of flight spectroscopy \cite{Avenhaus2009}, the linewidth of the idler photons was measured to be $0.20(4)$nm, in good agreement with the $0.14$nm linewidth expected from a KTP of length 50mm.

\begin{figure}[ht]
  \centering
  \includegraphics[width=0.45\textwidth]{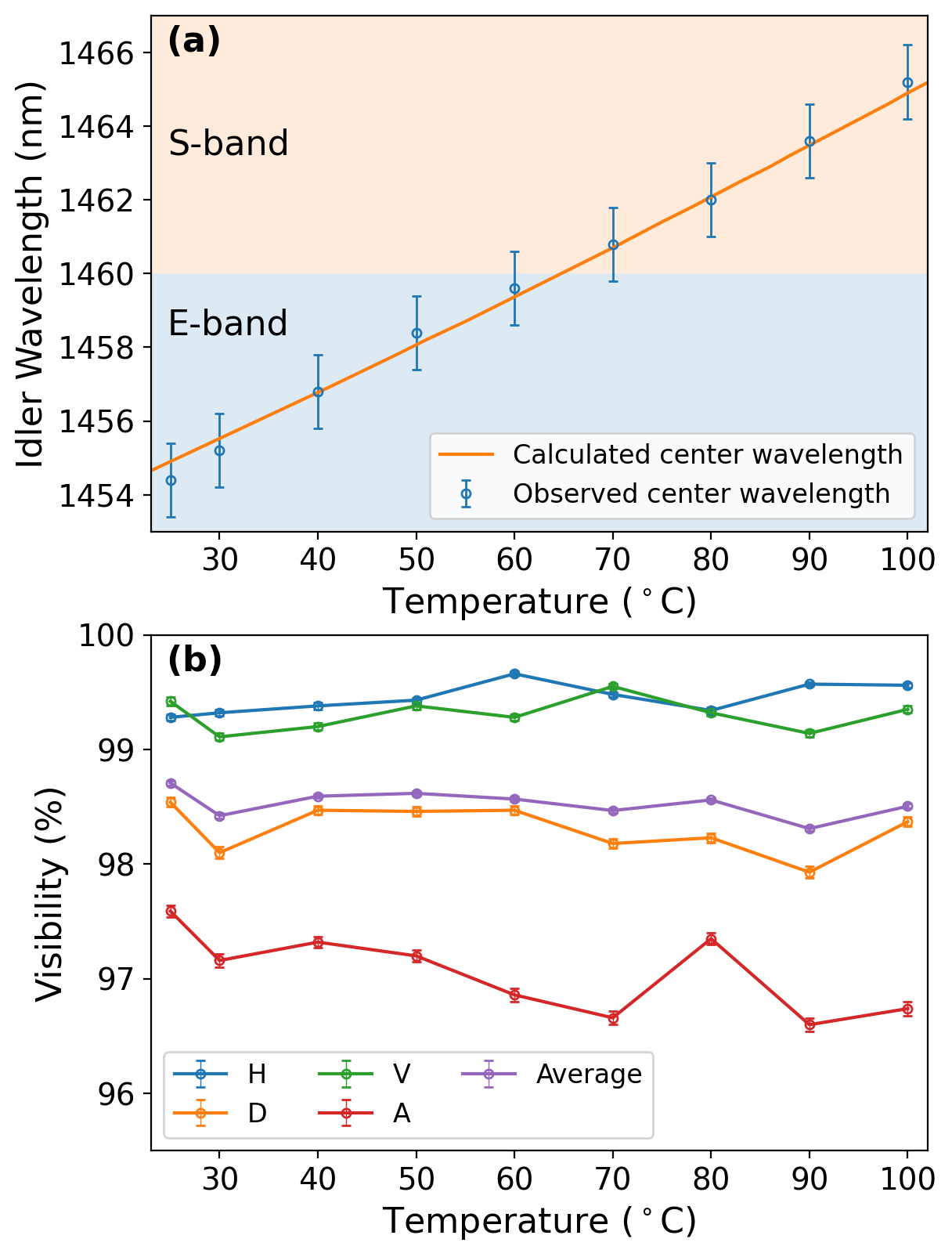}
\caption{(a) Graph of expected and observed center wavelength of the idler photons against temperature for NCPM in KTP with a pump wavelength of 405.75nm. The idler photons lie in the E- and S-bands. (b) Observed raw visibility of the source at various KTP temperatures. }
\label{fig:results}
\end{figure}

To verify the low temperature dependence of the output wavelengths in this NCPM scheme, the center wavelength of the idler photons was measured using a self-made grating spectrometer after heating the KTP to various temperatures. The entanglement visibility measurement was also repeated. The results are shown in Fig. \ref{fig:results}.

The center wavelength of the idler photons varied by (10.8$\pm\text{2)nm}$ over the temperature range of 25$^\circ$C to 100$^\circ$C. This corresponds to a center wavelength drift of 0.14(3)nm/$^\circ$C for the idler, in good agreement with the model. Due to the change in relative phase between the two downconversion paths as the output wavelengths change, the pump $\alpha$-BBO had to be tilted after significant temperature changes to restore the visibility in the D and A settings. After applying the tilt when necessary, the raw visibility averaged over the bases remain above 98$\%$ across the temperature range. 

\subsection{\label{section:distributed} Entanglement distribution over deployed fiber }

To test the feasibility of using the source for practical entanglement distribution, the idler photons were transmitted through 62km and 93km of deployed SMF28 fibers across Singapore \cite{Du2024} in separate experiments. The polarizations of the idler photons were analyzed after transmission through the deployed fibers. The idler photons were detected with the SNSPD, and the signal photons were detected at the source with the Si-APD. 

For a maximum fiber length of 93km and an idler linewidth of 0.2nm, the chromatic dispersion was estimated to be $\leq226$ps \cite{ITU2016, ITU2016a}, which was less than the timing jitter of the Si-APD. No significant chromatic dispersion effects were observed during the experiments, and a coincidence window of 1ns was used. $19$mW of pump power was used in both experiments. The results are summarized in Table \ref{tab:deployedresults}.



\begin{table}[ht]
\begin{ruledtabular}
\caption{\label{tab:deployedresults}Experiment results from entanglement distribution over deployed fiber. The idler photons were transmitted over deployed fiber. The signal photons were detected at the source.}
\begin{tabular}{>{\centering\arraybackslash}p{0.25\linewidth}>{\centering\arraybackslash}p{0.25\linewidth}>{\centering\arraybackslash}p{0.25\linewidth}>{\centering\arraybackslash}p{0.25\linewidth}>{\centering\arraybackslash}p{0.25\linewidth}}
    Length of deployed fiber (km) & Measured attenuation loss (dB) & Raw average visibility (\%) & CHSH S-parameter\\
    \hline
    62  & 24  & 98.2(1) & 2.78(4)  \\

    93   & 41  & 95.6(3) & 2.7(3)  \\
\end{tabular}
\end{ruledtabular}

\end{table}

The entanglement visibility remains above $95\%$ after transmission through the deployed fibers. The drop in visibility for the $93$km link is partially attributed to the low coincidence-to-accidentals ratio. After correcting for the accidentals, the average visibility for this link is $96.7(3)\%$. Depolarization effects due to polarization mode dispersion \cite{Zhang2024a, Lim2016} were not significant enough to cause large degradation in the entanglement visibility.

\section{\label{section:conclusion} Conclusion}
In this work, a source of highly nondegenerate polarization-entangled photon pairs based on NCPM SPDC in single-domain bulk KTP was demonstrated for the first time. The low temperature dependence of the output wavelengths was predicted and verified. Due to the small linewidth of the idler photons, chromatic and polarization mode dispersion effects were not significant, and high quality entanglement was shown to be preserved after transmission through deployed fiber in a metropolitan setting. These factors make photon-pair sources based on NCPM SPDC in KTP suitable for heterogeneous entanglement distribution with lower SWaP requirements. Additionally, the cost of NCPM sources can be much lower than that of conventional QPM sources - the 50mm single-domain KTP used in this work was an order of magnitude cheaper than commercial 30mm periodically poled KTPs. Other crystals might also be competitive in price and performance, and worth exploring for generating highly nondegenerate photon pairs in multi-channel entanglement distribution applications.

A challenge in using NCPM to generate idler photons in the O-band or C-band is to find a high-quality pump of the appropriate wavelength because NCPM only occurs for specific wavelength combinations in each type of crystal. This is especially true in temperature-insensitive cases such as in the case of KTP highlighted in this work. This challenge is made easier in crystals where temperature tuning is effective, which increases the range of output wavelengths available to each pump wavelength.

NCPM can be explored for generating photon pairs in waveguides where the phasematching conditions might be modified by waveguide dispersion. While more complicated, it allows for higher photon-pair generation rates and interesting structures like resonators to be achieved. 

\section{\label{section:acknowledgments} Acknowledgments}
This project is supported by the National Research Foundation, Singapore through the National Quantum Office, hosted in A*STAR, under its Centre for Quantum Technologies Funding Initiative (S24Q2d0009). We also acknowledge Netlink Trust and the National Quantum Safe Network for the provisioning of and the supporting access to the fiber network respectively.

\bibliography{Jabreflibrary}

\end{document}